\documentclass{aastex631}

\usepackage{amsmath}
\usepackage{subfigure}
\usepackage[misc]{ifsym} 
\usepackage{textcomp}

\usepackage{hyperref}

\shortauthors{Huang et al.}
\graphicspath{{./}{figures/}}

\begin{document}

\title{A solution to continuous RFI in narrowband radio SETI with FAST: The MultiBeam Point-source Scanning strategy}

\author[0000-0002-8719-3137]{Bo-Lun Huang}
\affiliation{Institute for Frontiers in Astronomy and Astrophysics, Beijing Normal University, Beijing 102206, China}
\affiliation{Department of Astronomy, Beijing Normal University, Beijing 100875, China}
\author[0000-0002-4683-5500]{Zhen-Zhao Tao}
\affiliation{Institute for Frontiers in Astronomy and Astrophysics, Beijing Normal University, Beijing 102206, China}
\affiliation{Department of Astronomy, Beijing Normal University, Beijing 100875, China}
\affiliation{Institute for Astronomical Science, Dezhou University, Dezhou 253023, China}
\author[0000-0002-3363-9965]{Tong-Jie Zhang\href{mailto:tjzhang@bnu.edu.cn}{\textrm{\Letter}}}
\affiliation{Institute for Frontiers in Astronomy and Astrophysics, Beijing Normal University, Beijing 102206, China}
\affiliation{Department of Astronomy, Beijing Normal University, Beijing 100875, China}



\begin{abstract}

Narrowband radio search for extraterrestrial intelligence (SETI) in the 21st century suffers severely from radio frequency interference (RFI), resulting in a high number of false positives, and it could be the major reason why we have not yet received any messages from space. We thereby propose a novel observation strategy, called MultiBeam Point-source Scanning (MBPS), to revolutionize the way RFI is identified in narrowband radio SETI and provide a prominent solution to the current situation. The MBPS strategy is a simple yet powerful method that sequentially scans over the target star with different beams of a telescope, hence creating real-time references in the time domain for cross-verification, thus potentially identifying all continuous RFI with a level of certainty never achieved in any previous attempts. By applying the MBPS strategy during the observation of TRAPPIST-1 with the FAST telescope, we successfully identified all 6972 received signals as RFI using the solid criteria introduced by the MBPS strategy. Therefore we present the MBPS strategy as a promising tool that should bring us much closer to the first discovery of a genuine galactic greeting.

\end{abstract}



\section{Introduction}

For millennia, humans have gazed at the stars and maybe wondered if we are also being gazed at. With generations of effort poured into the field of astronomy, SETI is now technologically capable to explore the old question raised by our stone age ancestors.
Recent SETI research has predominately focused on narrowband ($\sim $ Hz) radio technosignatures due to their distinctiveness from natural radio emissions, and efficiency in transmission, thus it is assumed that advanced civilizations could also choose it for communication \citep{sheikh2023green,tarter2001search,li2020opportunities}. The narrowband feature($\sim $ Hz) of a hypothetical signal is sufficient to filter out known astrophysical emissions. However, identifying radio frequency interference (RFI) remains a major challenge for SETI. The potential RFI sources include wireless communication, television, radar, artificial satellite, instrumental RFI, etc \citep{zhang2020rfi}. 

\vspace{\baselineskip}

In order to identify RFI, previous SETI studies have developed strategies such as the ON-OFF method for single-dish radio telescopes, where the telescope points to the target star for a few minutes and then points at a reference location that is at least six beamwidths away from the primary target \citep{enriquez2017breakthrough}. The target star is then the ‘ON’ source and the reference location is the ‘OFF’ source. An authentic extraterrestrial intelligence (ETI) signal sent from the target star would only appear in the ON observations and must not be in the OFF ones since they are sufficiently separated in the sky  \citep{nageneral}. Later, the MultiBeam Coincidence Matching (MBCM) strategy is proposed which utilizes The five-hundred-meter aperture spherical radio telescope (FAST)’s L-band 19-beam receiver \citep{nan2011five,li2018fast}. The central beam points right at the target, and the outermost six beams act as the ‘OFF’ observations so that the ‘ON’ and ‘OFF’ observations are taken simultaneously \citep{zhang2020first}. Based on MBCM, in order to maximize efficiency, a blind search mode is designed for MBCM, allowing any of the 19 beams to be considered as the 'ON' source for RFI removal\citep{luan}. Although no technosignatures have yet been confirmed, the aforementioned strategies have achieved significant sensitivity in terms of identifying RFI. However, some ground-based RFI like those that originated from crystal oscillators of a telescope often show similar characteristics to what we expected as a signal from ETI, such as the drifting frequency, and bandwidth \citep{tao}. \textbf{Continuous RFI can contaminate only one or two adjacent beams, passing through even the RFI removal pipeline of the MBCM, which results in an increase in the rate of false positives.}

\vspace{\baselineskip}

\begin{figure}[htb]
\centering
\subfigure[]{
\includegraphics[width=0.73\linewidth]{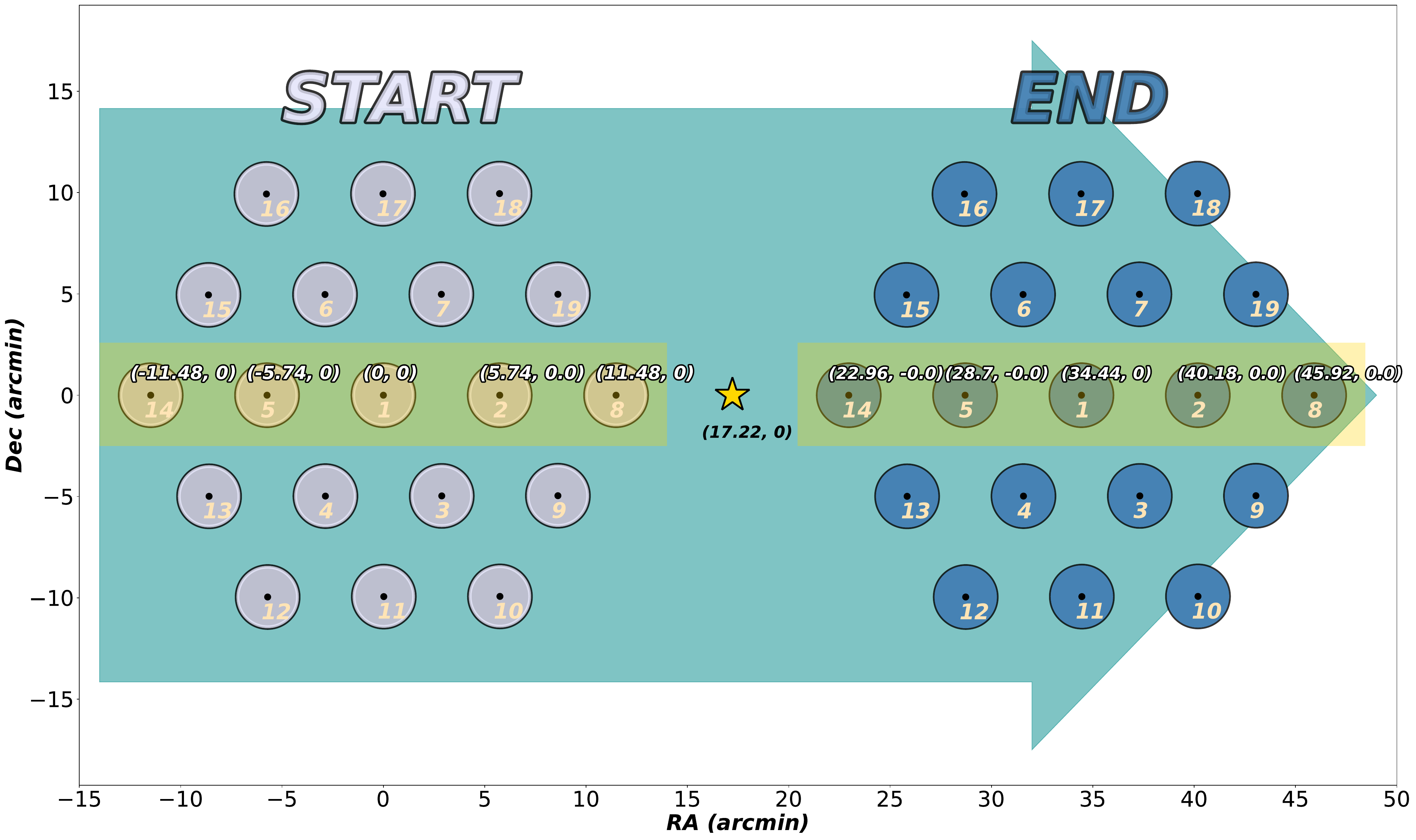}
}
\subfigure[]{
\includegraphics[width=0.73\linewidth]{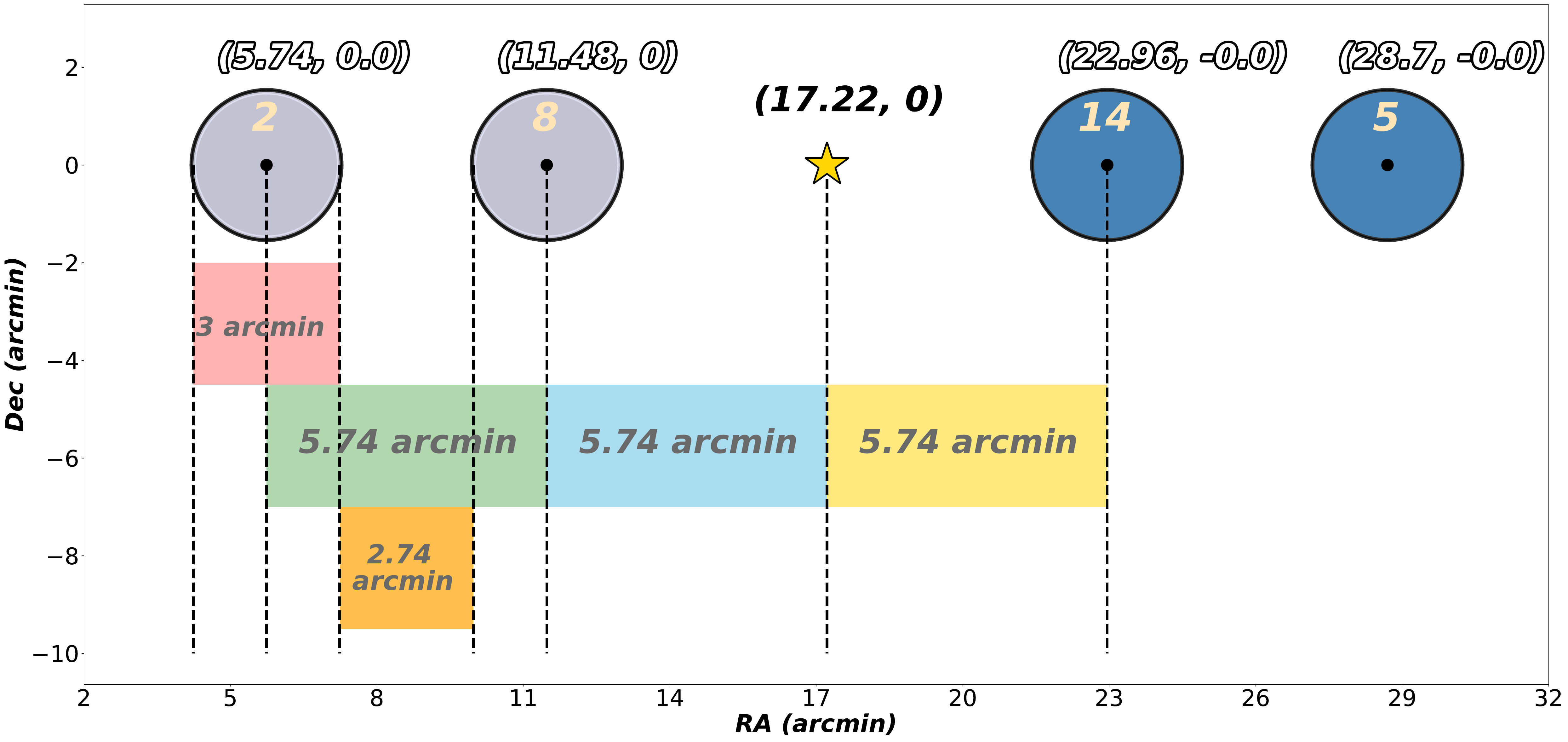}
}

\caption{(a) Relative positions for the 19 beams of FAST with rotation angle = 0° at the starting point of the MBPS observation are shown with grey circles, beam 1 is placed at the origin, and the five yellow-shaded beams are the On-Target Beams where the diameter of each circle is the half-power beamwidth (HPBW) of each beam. Positions of the blue circles represent the relative coordinates when MBPS observation ends. The star represents the position of the target which is placed at (17.22, 0) relative to beam 1. The cyan arrow indicates the direction of the telescope slew. (b) A zoom-in plot of (a) around the target, length of the red, green, and orange shaded regions in the x-axis represent the beam's HPBW, angular separation between adjacent beams (ASAB), and angular separation between HPBWs of adjacent beams respectively. }
\label{fig:19beam}
\end{figure}

In this paper, we propose a new observation strategy called MultiBeam Point-source Scanning (MBPS), aiming to solve the trouble brought by continuous RFI. In this work, the MBPS is designed based on the relative positions of the 19 beams of the FAST’s 19-beam receiver \citep{jiang2020fundamental} as can be seen from Fig. \ref{fig:19beam}(a). The core idea is that a sky-localized signal would show similar but not identical features as different beams scan over it. Therefore enabling us to make some of the strictest rules on ETI confirmation as well as RFI identification. Unlike most narrowband SETI surveys where the telescopes remain still relative to the target during observations. MBPS strategy requires the telescope to constantly slew during an observation as shown in Fig. \ref{fig:19beam}(a), thus effectively providing real-time reference locations to be compared with for each beam. In addition, we conduct a re-observation of TRAPPIST-1 with FAST's 19-beam receiver using the MBPS observation strategy and aim to test the feasibility of the new strategy.

\clearpage

\section{Methods}

\subsection{MBPS observation strategy}

The MBPS observation requires the rotation angle of the FAST's L-band 19-beam receiver to be 0 degrees so that beams 8, 2, 1, 5, and 14 are at the same declination as shown in Fig. \ref{fig:19beam}(a), here we define the yellow-shaded five beams as the On-Target Beams (hereafter OTBs) as they are the only beams that scan right on the target during an MBPS observation. The MBPS observation strategy operates in the following steps:

\vspace{\baselineskip}

1, Beam 1 (the central beam) starts at the same declination as the target, and -17.22 arcmin (3 times the angular separation of adjacent beams(ASAB)) in right ascension away from the target, which means beam 8 (the rightmost beam) is now exactly 5.74 arcmin (angular separation of adjacent beams) from the target as shown in Fig. \ref{fig:19beam}(b).

2, The telescope scans at the same declination with a constant slew rate in the positive right ascension direction (cyan arrow in Fig. \ref{fig:19beam}(a)).

3, The observation ends when beam 1 is exactly at +17.22 arcmin in right ascension away from the target so that beam 14 (the leftmost beam) is now exactly 5.74 arcmin from the target as shown in Fig. \ref{fig:19beam}(b).

\vspace{\baselineskip}

\subsection{Cross-verification for ETI signals}

The most important feature of the MBPS strategy is that it enables us to cross-verify whether a signal is sky-localized by inspecting multiple newly introduced parameters. We then explore the variations of the new parameters which we benefit from the fact that the telescope is constantly slewing during an MBPS observation. Unlike any targeted SETI observations that have been conducted with FAST, the MBPS strategy provides solid reasoning for any potential ETI signals in its verification process. 

\subsubsection{On-Target Window}
As the telescope scans from left to right in Fig. \ref{fig:19beam}(a), The target subsequently enters the half-power beam width (HPBW) circles of the five OTBs(Beams 8, 2, 1, 5, and 14) in a strict order. Therefore, we can predict what would a real ETI signal look like in dynamic spectrum plots (time-frequency plots) of the OTBs. If we have the information on the total observation time and the slew rate of the telescope. We start by first calculating the time stamps when the target enters and exits the HPBW circles of each of the five OTBs (Beam 8, 2, 1, 5, 14) from the start of the observation, we define the time windows between the target entering and exiting an OTB’s HPBW circle as the On-Target Window (hereafter OTW). Since the telescope slews at a constant rate and the HPBW for all 19 beams are 3 arcmin (the HPBW can vary slightly depending on the frequency of received signals, for simplicity here we adopt the HPBW at 1300 MHz which is approximately 3 arcmin ) \citep{dong2013beam,jiang2020fundamental}, the length of an OTW can be computed: HPBW/slew rate = 3 arcmin/ slew rate. Between each OTW, there are four gap windows, which are defined as the time windows between two consecutive OTWs, the gap windows start at the exiting time stamps of an OTW, and end at the starting time stamps of the next OTW and the length of gap window can be calculated as (ABAS-HPBW)/slew rate = 2.74 arcmin/slew rate. A continuous ETI signal originating at the target will show up in multiple OTWs but centers at different frequencies depending on its drift rate, therefore enabling us to cross-verify the presence of an ETI signal by examining the consistency of the signal in different OTWs.

\subsubsection{Maximum intensity of an ETI signal}

Furthermore, signals from the target star would reach the maximum intensity when the target star is exactly at the center of a beam (intensity maximum occurs at the mid-point in an OTW). 

\subsubsection{Intensity variation}

\begin{figure}[htb]
\centering
\subfigure[]{
\includegraphics[width=0.4\linewidth]{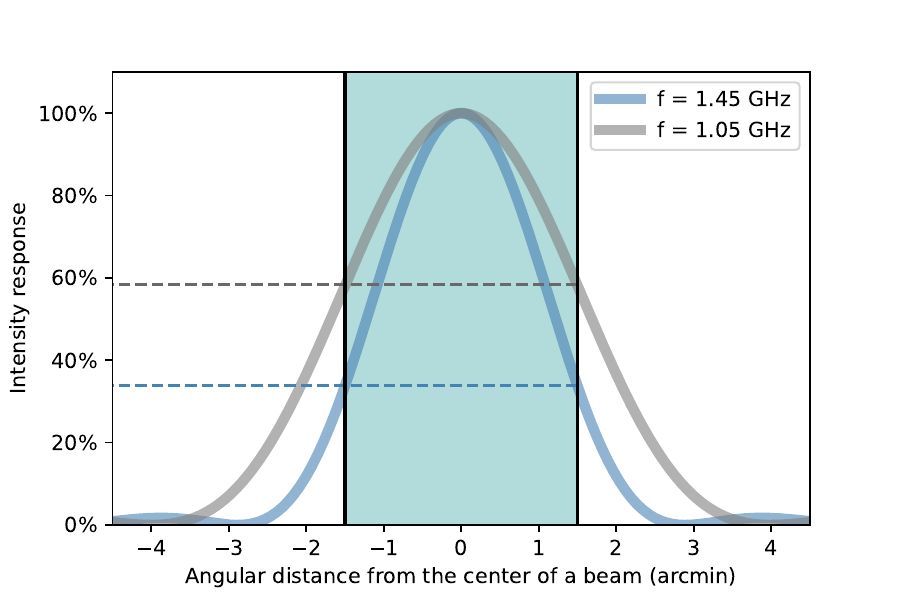}
}
\subfigure[]{
\includegraphics[width=0.4\linewidth]{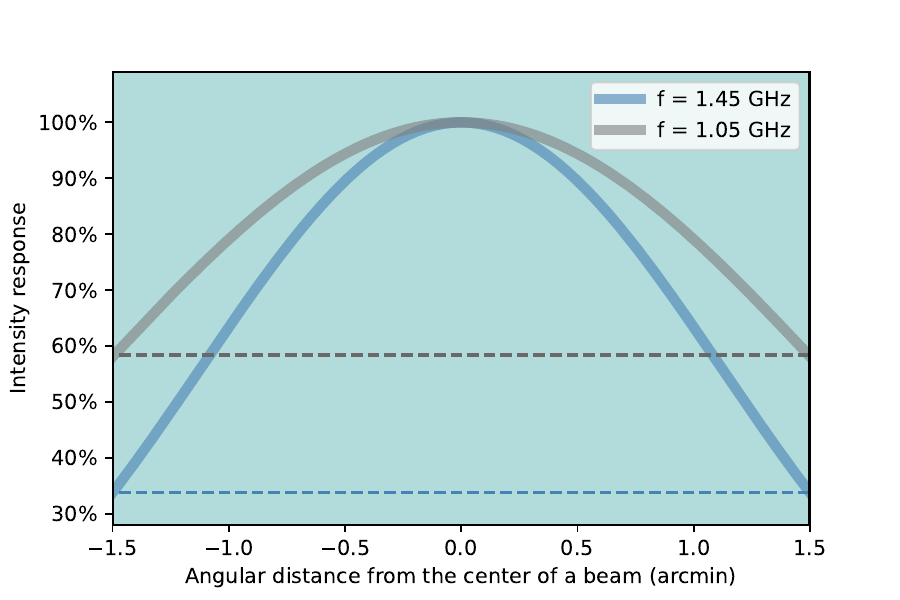}
}

\caption{(a) The intensity response of a beam in the 19-beam receiver of FAST, the origin represents the center of a beam. (b) The predicted intensity variation of a sky-localized signal inside an OTW. The gray line is the response when the received signal is at a frequency of 1.05 GHz and the blue line is the response when the signal is at 1.45 GHz (the frequency band we use for SETI observation at FAST is 1.05-1.45 GHz). }
\label{fig:resp}
\end{figure}

For FAST, the intensity variation of a sky-localized signal inside an OTW can be roughly represented by the beam response of a uniformly illuminated circular aperture telescope which is given by 

\begin{align}
I(\theta) &= I_{0} \left(\frac{2J_{1}(k a \sin(\theta))}{k a \sin(\theta)}\right)^2 ,
\end{align}

where $I(\theta)$ and $I_{0}$ stand for intensity response at $\theta$ from the center of a beam and the intensity at the center of a beam, $J_{1}$ is the Bessel function of the first kind with the order of one, $k$ is the wavenumber of the frequency that is being received, and $a$ is the radius of the telescope aperture. Rearrange and we get 

\begin{align}
\frac{I(\theta)}{I_{0}} &=\left(\frac{2J_{1}(k a \sin(\theta))}{k a \sin(\theta)}\right)^2 ,
\end{align}

where the left-hand side is the ratio of the intensity response at $\theta$ and the maximum response, thus we could plot the ${I(\theta)}/{I_{0}}$ against $\theta$ as shown in Fig. \ref{fig:resp}. According to \citet{jiang2020fundamental}, the intensity responses are slightly different for each of the 19 beams, and the performance of FAST's 19-beam receiver shows a combination of a theoretical cosine-tapered illumination and a theoretical uniform illumination. However, the overall pattern remains the same that is the response of a signal decreases significantly as the signal leaves the center of a beam. As we can see from Fig. \ref{fig:resp}, the responses of a beam are different at different frequencies, which means once a signal is determined to be a potential ETI signal by its OTW and frequency, we need to further investigate if its intensity variation follows the theoretical values for that specific frequency.

\subsubsection{Simulated ETI signals}

\begin{figure}[htb]
\centering
\subfigure[]{
\includegraphics[width=0.3\linewidth]{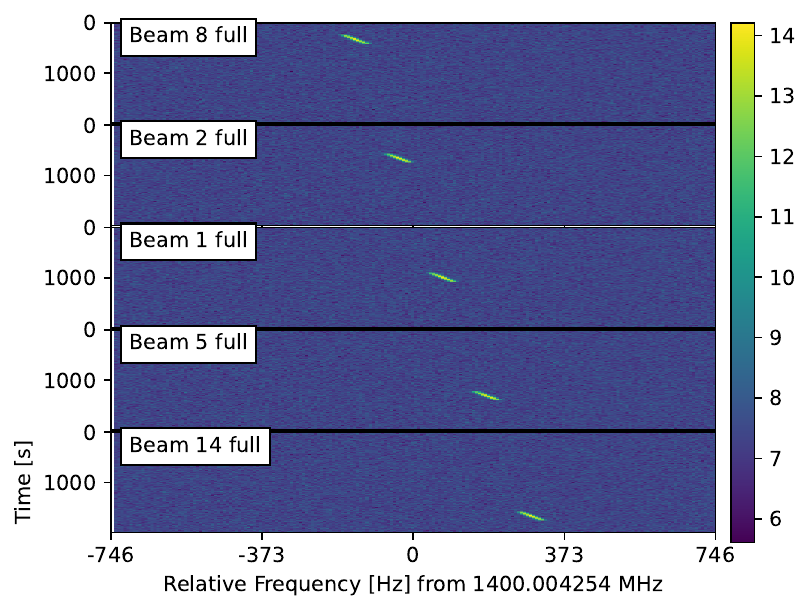}
}
\subfigure[]{
\includegraphics[width=0.3\linewidth]{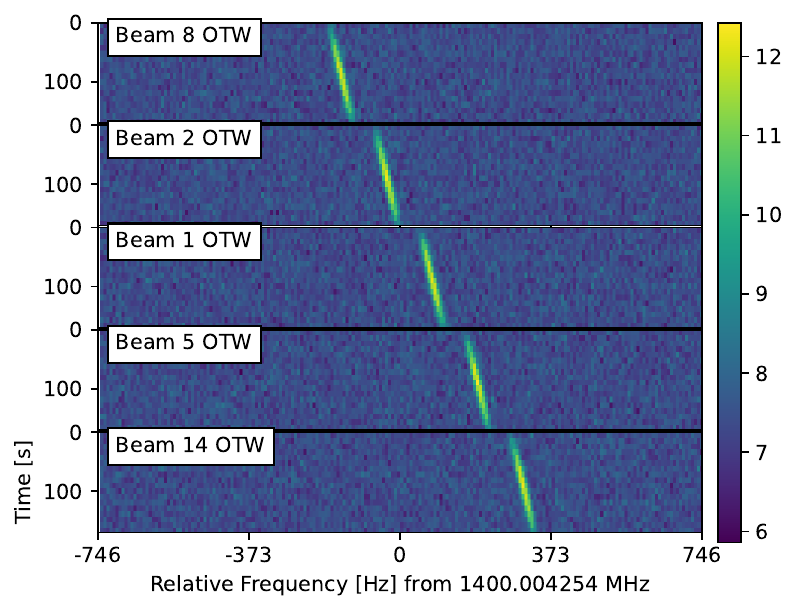}
}
\subfigure[]{
\includegraphics[width=0.3\linewidth]{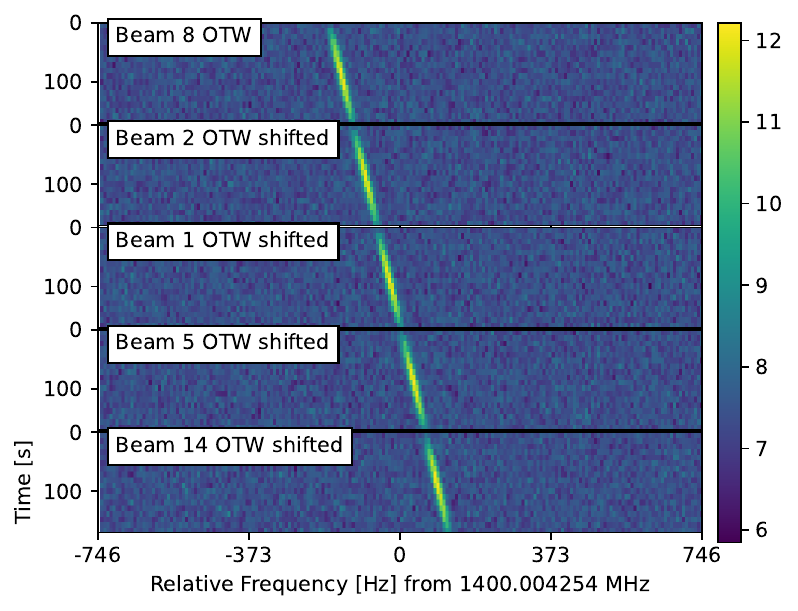}
}

\caption{(a) Waterfall plots for a simulated ETI signal in the five OTBs cross the full observation time. (b) Waterfall plots for a simulated ETI signal in the five OTBs during their five OTWs, each with a length of 181 seconds. (c) Waterfall plots for a simulated ETI signal in the five OTBs during their OTWs with corrected frequencies. The colored bar indicates the power with arbitrary units. }
\label{fig:uncut}
\end{figure}

We use the setigen python package \citep{enriquez2019turboseti,brzycki2022setigen} to demonstrate how a potential ETI signal that lasts longer than the total observation time of 2074 seconds and a constant telescope slew rate of 1 arcsec/s, would appear in the waterfall plots under the MBPS strategy. The hypothetical ETI signal has a constant drift rate of 0.33 Hz/s and a bandwidth of 15 Hz, as can be seen in Fig. \ref{fig:uncut}(a), the signal only occupies a small portion of the total observation time in each beam with different central frequencies. However, if we only plot the five OTWs of beams 8, 2, 1, 5, and 14, and discard those outside the OTWs for each beam as shown in Fig. \ref{fig:uncut}(b), we could clearly see that signals in the OTWs are consistent in drift rate but with small shifts in frequency. The small shift in frequency comes from the gaps between OTWs, that is instead of being in the HPBW of an OTB, the target is now between the HPBW of two adjacent beams (the orange-shaded region in Fig. \ref{fig:19beam}(b)). The shifts can be corrected by calculating the amount of frequency drift during the gaps, and subtracting it from the observed frequencies of beams 2, 1, 5, 14, hence we obtain Fig. \ref{fig:uncut}(c), and therefore, the signal is cross-verified with all five OTWs. Furthermore, the simulated signal exhibits a characteristic of being strong in the middle and weaker as it extends to edges due to the sensitivity of the beam decreasing as the target moves away from the center of the beam.
\clearpage

\vspace{\baselineskip}

\subsection{RFI identification}

\begin{figure}[htb]
\centering
\subfigure[]{
\includegraphics[width=0.4\linewidth]{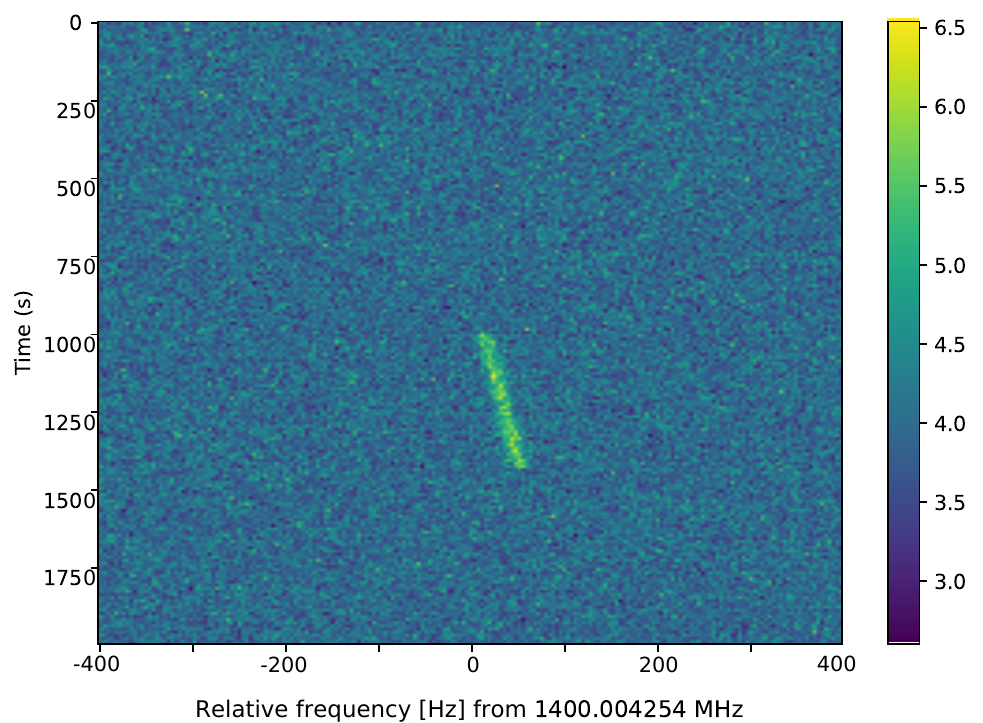}
}
\subfigure[]{
\includegraphics[width=0.4\linewidth]{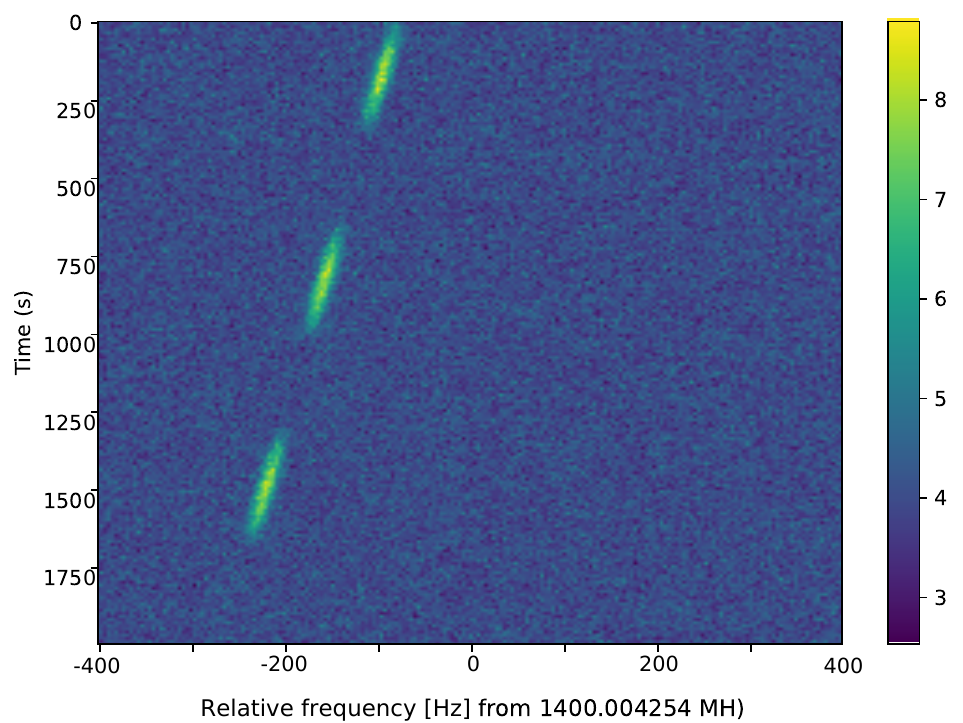}
}
\subfigure[]{
\includegraphics[width=0.4\linewidth]{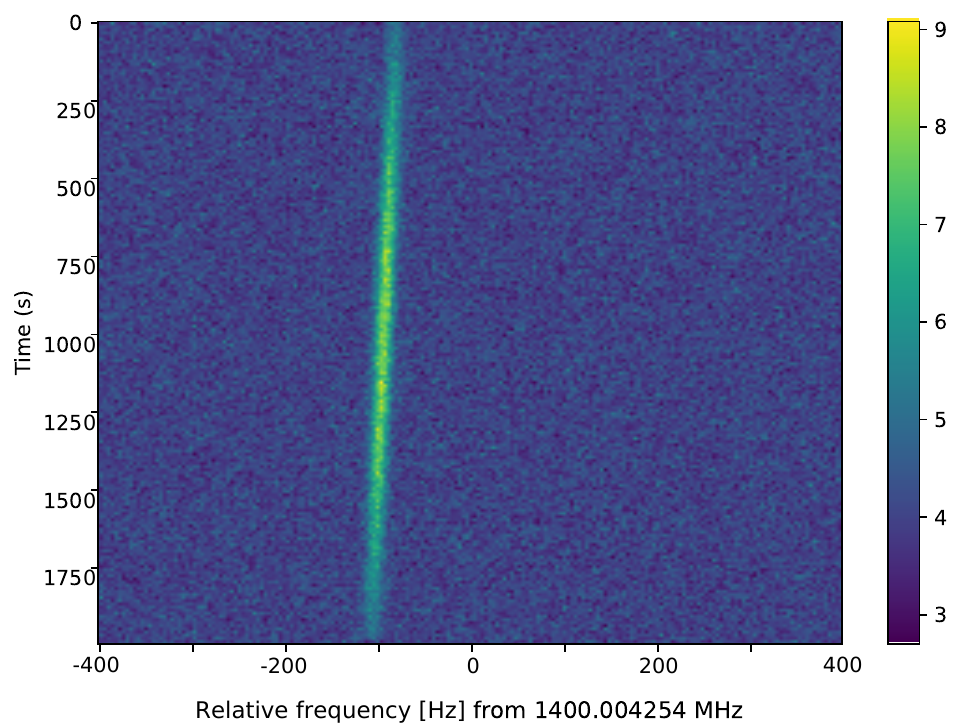}
}
\subfigure[]{
\includegraphics[width=0.4\linewidth]{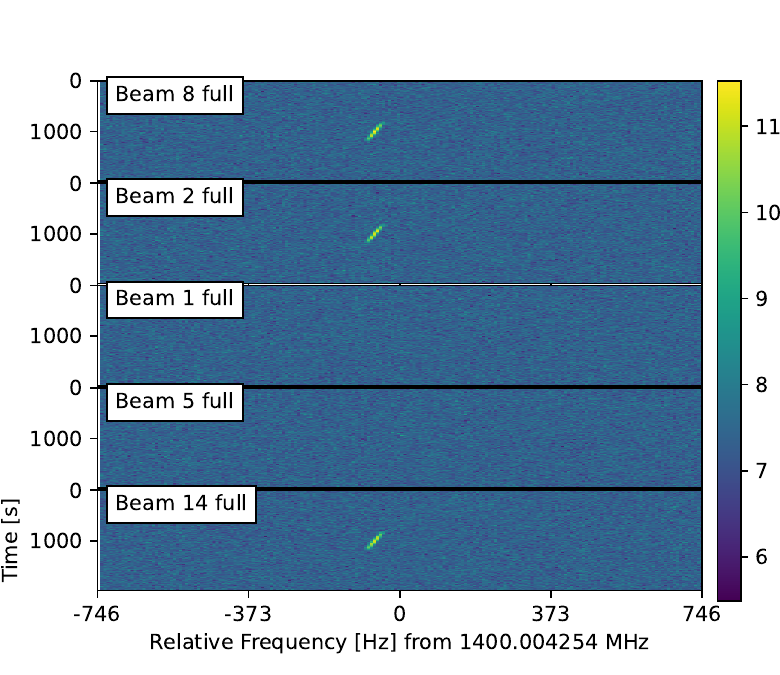}
}

\caption{Four examples of simulated RFI that could be identified by MBPS. The colored bar indicates the power with arbitrary units. (a) signal that appears for a fraction of the observation, but with constant intensity. (b) intermittent signal that is across a long duration. (c) signal that is continuous throughout the observation. (d) signal that appears simultaneously at different beams but with the same central frequency.}
\label{fig:RFI}
\end{figure}

An instrumental RFI, instead of occupying only a portion of observation time and being consistent in the five OTWs, would show up in multiple random beams at the same time or occupy one or more beams for the entire observation time because ground-based and instrumental RFI are not originated in a particular sky position, movements of the telescope pointing does not determine which beam would be detecting them at a specific time. Satellites like GPS satellites with large beamwidth can be detected by the 19-beam receiver of FAST \citep{tao}. Therefore, such a signal should be visible in multiple beams with the same central frequencies and is immediately marked as RFI according to the MBPS strategy. Satellites with a small beamwidth that is comparable to or smaller than the HPBW of FAST can be identified by searching in the FAST satellite RFI database \citep{wang2021satellite}. The MBPS strategy can also be a tool to identify RFI from satellites by the intensity variations in different OTWs. Fig. \ref{fig:RFI} shows four simulated examples of possible RFI. Assuming the telescope starts at time = 0s, the rules of MBPS strategy to identify RFI are as follows: 

1, signals that span the entire observation time in any of the beams are immediately marked as RFI as shown in Fig. \ref{fig:RFI}(c). 

2, An ETI signal from the target would appear only in the OTWs as shown in Fig. \ref{fig:uncut}(a), or if the signal is powerful enough, it may extend to both edges by a few seconds but show a dramatic intensity falling as it extends as shown in Fig. \ref{fig:uncut}(b)(c). However, signals with a duration much greater than OTW or their intensity stays constant can be marked as RFI as shown in Fig. \ref{fig:RFI}(a)(b), because the fact that the pointing of the telescope would have pointed away from the target, and the target is no longer within the HPBW circle of that beam. 

3, Signals that are detected at the same time with the same central frequency in multiple beams are marked as RFI as shown in Fig. \ref{fig:RFI}(d), because the angular separation of beams and the movement of the telescope are adequate to ensure that detected intensities of a space-originated signal would differ significantly at the same time in even adjacent beams.

\vspace{\baselineskip}

\section{Results} \label{sec:Results}

\subsection{Observation of TRAPPIST-1} \label{sec:t1}

TRAPPIST-1 is a nearby star ($\sim$12 pc) known to host 7 planets \citep{grimm2018nature,luger2017seven} which makes it a perfect target for SETI survey that focuses on nearby stars \citep{lustig2019detectability}. In 2021, a 20-minute targeted observation was conducted to TRAPPIST-1 \citep{tao}, though no ETI was confirmed. The new observation of TRAPPIST-1 is carried out by FAST’s 19-beam receiver, starts at MJD 60077.02014, and ends at 60077.04414, lasting a total of 2074 seconds. The data is acquired through the L-band 19-beam receiver, operating with the SETI backend mode, spanning a frequency range from 1.05 to 1.45 GHz. The frequency resolution is approximately 7.5 Hz, while the time resolution is about 10 seconds. The data is then processed by TurboSETI, a software package that employs the \textbf{”Taylor tree deDoppler”} algorithm to search for narrowband drifting signals \citep[etc.]{taylor1974sensitive,siemion20131,enriquez2017breakthrough,gajjar2021breakthrough}. It searches for signals with drift rates within the maximum drift rate (MDR) and signal-to-noise ratio (SNR) above a preset threshold. We continue to use the SNR threshold of 10 that has been widely used in surveys targeting nearby stars \citep{price2020breakthrough}. We calculated the maximum drift rate of a hypothetical transmitter that is located at the equator of TRAPPIST-1c (the second innermost planet in the TRAPPIST-1 system, and it is outside the habitable zone of TRAPPIST-1 system) which is \textbf{10.60 Hz/s \citep{li2022drift} and the MDR is thus set to 11 Hz/s} in TurboSETI. The On-Target windows of the five OTBs can be calculated prior to the observation as shown in Table. \ref{tab:otw}, based on that the slew rate is kept constant at approximately 1 arcsec/s, and a total observing time of 2074 seconds.

\begin{longtable}{c c c}%
\caption{On-Target Windows of the five OTBs}
\label{tab:otw}
\\
\hline%
\hline%
Beam Number & On-Target Window (time elapsed from the start of observation in second) & OTW length (second)\\
\hline%
\endhead%
\hline%
\endfoot%
\hline%
\endlastfoot%

Beam 8 & 255 - 436 & 181 \\

Beam 2 & 601 - 782 & 181 \\

Beam 1 & 947 - 1128 & 181 \\

Beam 5 & 1293 - 1474 & 181 \\

Beam 14 & 1639 - 1820 & 181 \\

\end{longtable}

\subsection{Data analysis} \label{sec:t2}

During the 2074 seconds of observation, $\sim$3.7 T of observation data is generated. 
We then run the TurboSETI software to search for narrowband drifting signals only in the five OTBs. A total number of 6972 hits\footnote{Hits are signals that fulfil the conditions of SNR $>$ 10 and MDR within $\pm 11$ Hz/s. were found. The sensitivity of the drift rate of this observation is calculated to be $0.0036 Hz/s$ which corresponds to an acceleration of a mere $0.00077 m/s^2$ which is two orders of magnitudes less than the the acceleration caused only by the rotation and orbit of Earth. Thus, we remove hits with zero drift rate which account for 85.5 \% of the total hits.} Hits with zero drift rate are discarded because a zero drift rate often means that there is no relative acceleration between the source of the hit and the telescope and thus most likely to originate on-ground \citep{li2022drift}. We then apply the MBPS RFI identification strategy through visual inspection as mentioned in the Methods section \textbf{(Section 2)} to the remaining 1011 hits. Not a single signal passes the MBPS strategy, out of the 1011 hits, 988 hits show a continuous signal throughout the entire observation; 14 hits consist of signals that only appear for a fraction of the observation but with constant intensity; the remaining 9 hits though exhibit intensity variation and last for only a small portion of the observation but show up in multiple beams at the same central frequencies.

\begin{figure}[htb]
\centering
\subfigure[]{
\includegraphics[width=0.4\linewidth]{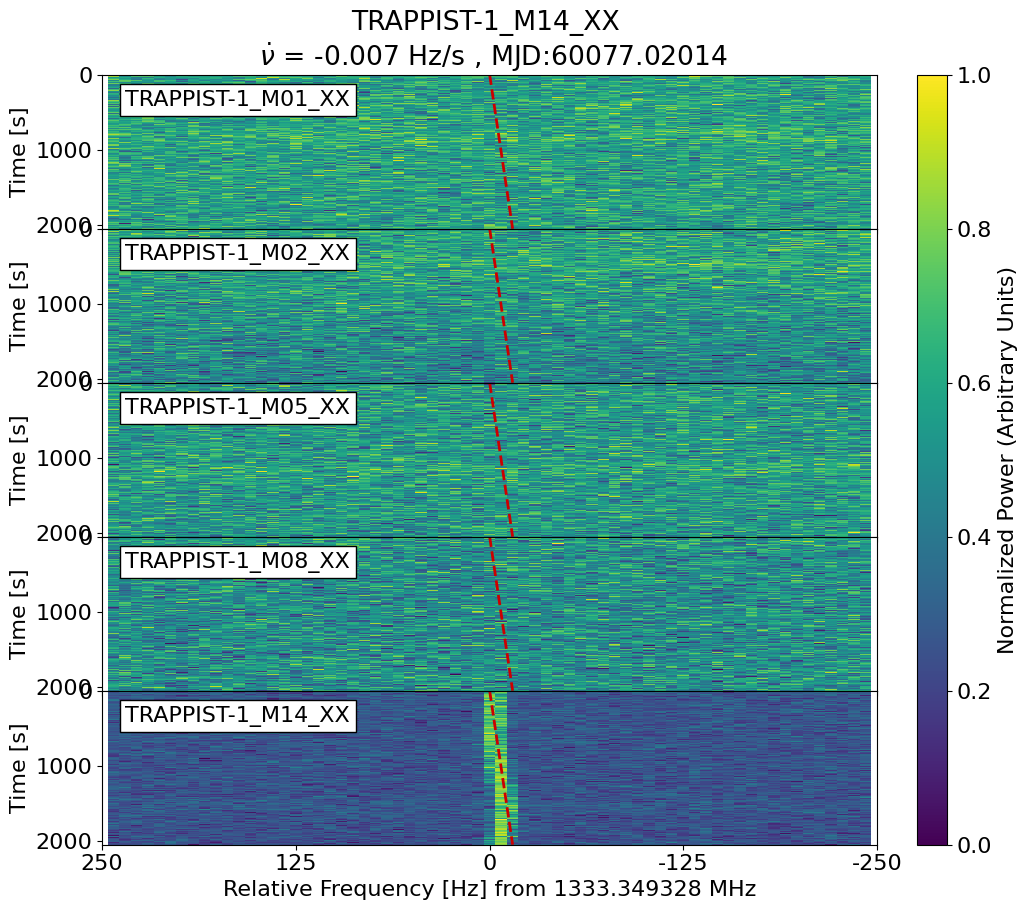}
}
\subfigure[]{
\includegraphics[width=0.4\linewidth]{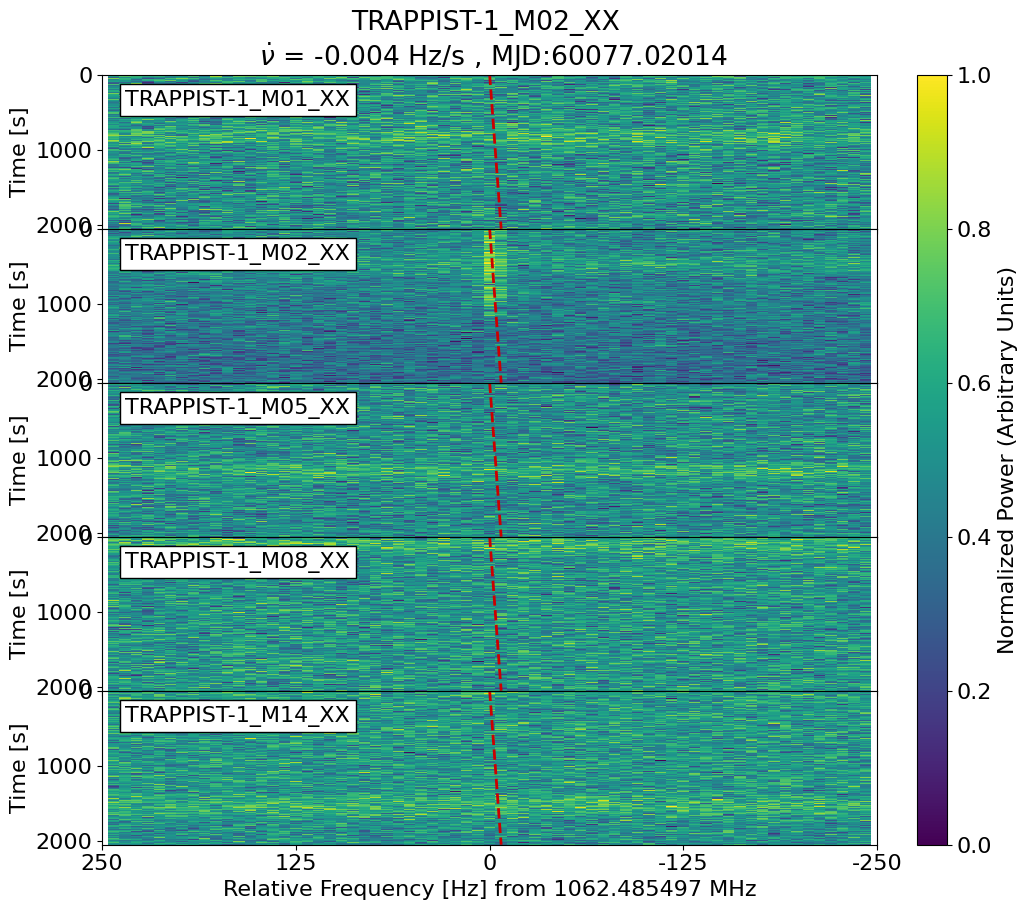}
}
\subfigure[]{
\includegraphics[width=0.4\linewidth]{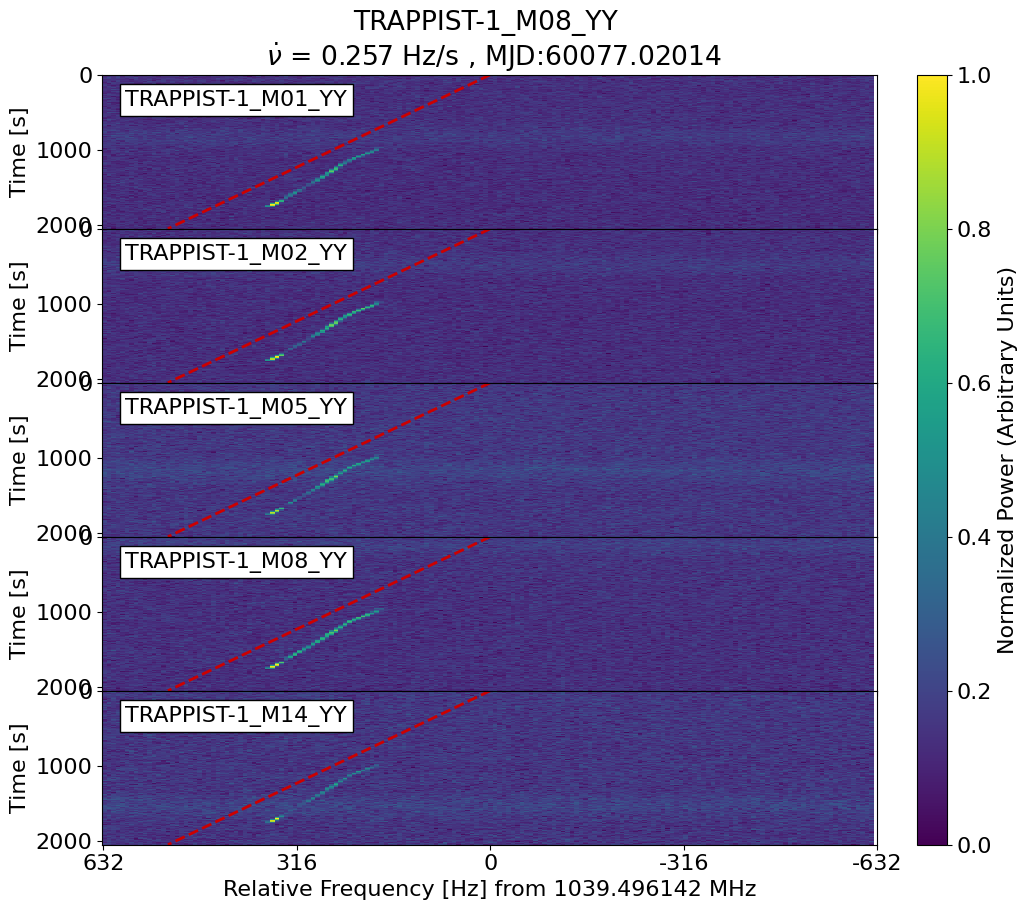}
}
\subfigure[]{
\includegraphics[width=0.4\linewidth]{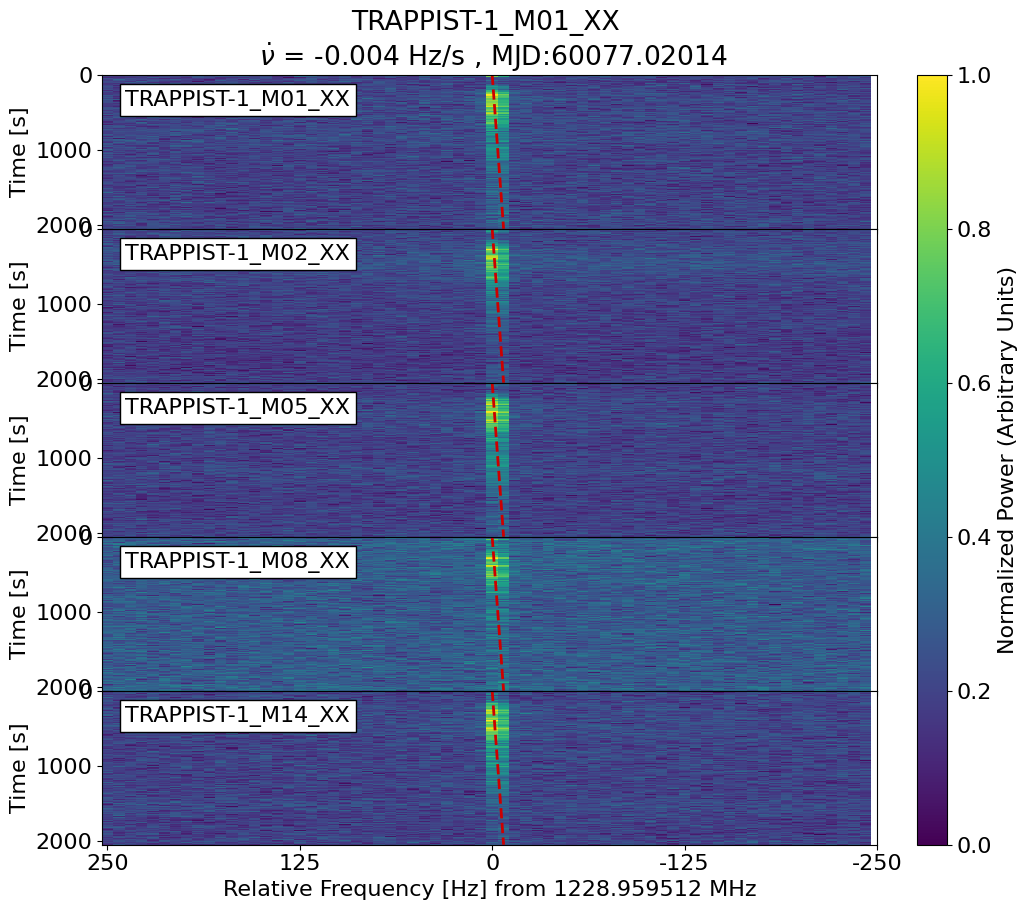}
}

\caption{Examples of RFI in the observation of TRAPPIST-1. (a) The signal is visible throughout the full observation time in beam 14 but not in any other beam. (b) The signal only appears in beam 2 for a fraction of observation but shows constant intensity. (c) The signal occupies a fraction of observation and shows variation in intensity but is visible in all OTBs and exhibits the same central frequency. (d) The signal shows an intensity falling during the entire observation in all OTBs simultaneously.  }
\label{fig:realrfi}
\end{figure}

\vspace{\baselineskip}

\subsection{Ground based RFI}

Instrumental RFI of FAST often come from crystal oscillators used in the digital back-end of the 19-beam receiver. In previous strategies, Instrumental RFI can be recognized by checking if their central frequencies fall near a known instrumental RFI region. However, it does not rule out the possibility that an ETI signal is coincidentally sent at those frequencies. As can be seen in Fig. \ref{fig:realrfi}(a), the signal in beam 14 is located around 1333.349328 MHz which is one of the known instrumental RFI frequencies caused by clock oscillators \citep{tao}. The MBPS strategy allows us to put a two-step verification on it by inspecting the signal's duration, a signal that spans the entire observation time is marked as RFI regardless of its central frequency. Sky-localized signals detected by the MBPS strategy would show a significant rise and drop when the target enters and leaves the HPBW circle of a beam. Therefore, signals with a constant intensity for a time longer than the OTW are marked as RFI as shown in Fig. \ref{fig:realrfi}(b), the signal lasts for more than 1000 seconds which is multiple times the length of an OTW while exhibits no significant variation in intensity. As for Fig. \ref{fig:realrfi}(c), the signal is visible in all five OTBs and shows intensity variation, however, the central frequencies in the five beams are not consistent with the movement of the telescope. The signal presented in Fig. \ref{fig:realrfi}(c)  is therefore identified as RFI by the MBPS strategy.

\vspace{\baselineskip}

\subsection{Possible satellite RFI}

 Fig. \ref{fig:realrfi}(d) shows a potential satellite RFI because the signal is visible in all five beams with the same central frequency and all of them exhibit a slow decrease in intensity during the 2074 seconds of observation which could suggest that a satellite is passing by the pointing of the telescope, the intensity decreases due to the increasing distance between the satellite and the telescope pointing. 

\subsection{Higher slew rate of telescope}

\begin{figure}[htb]
\centering
\subfigure[]{
\includegraphics[width=0.4\linewidth]{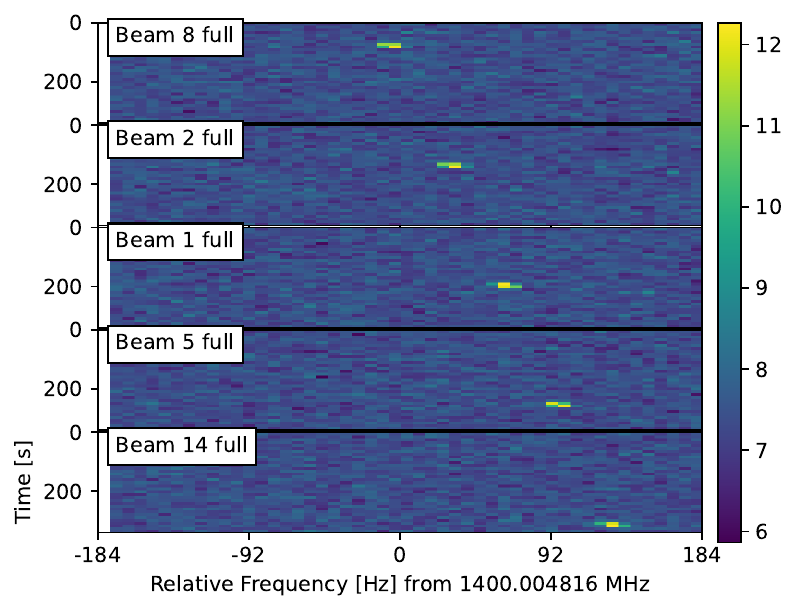}
}
\subfigure[]{
\includegraphics[width=0.4\linewidth]{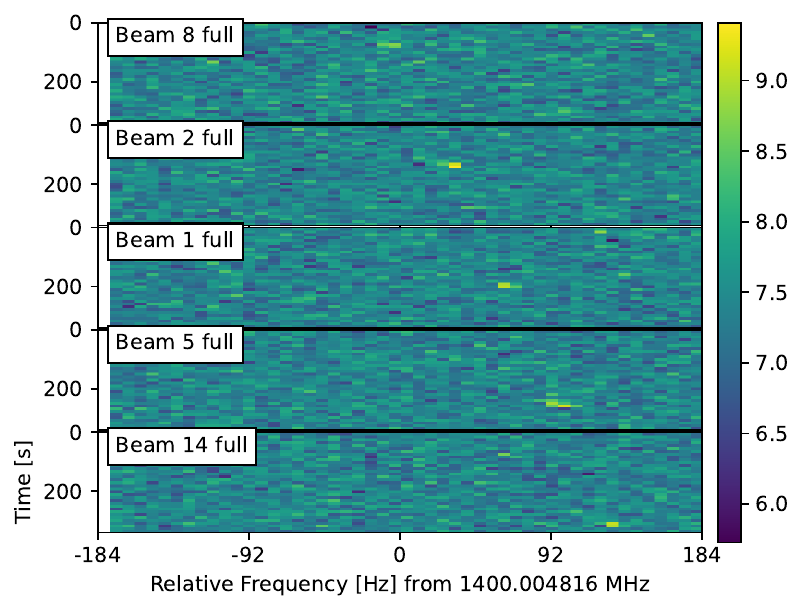}
}
\subfigure[]{
\includegraphics[width=0.4\linewidth]{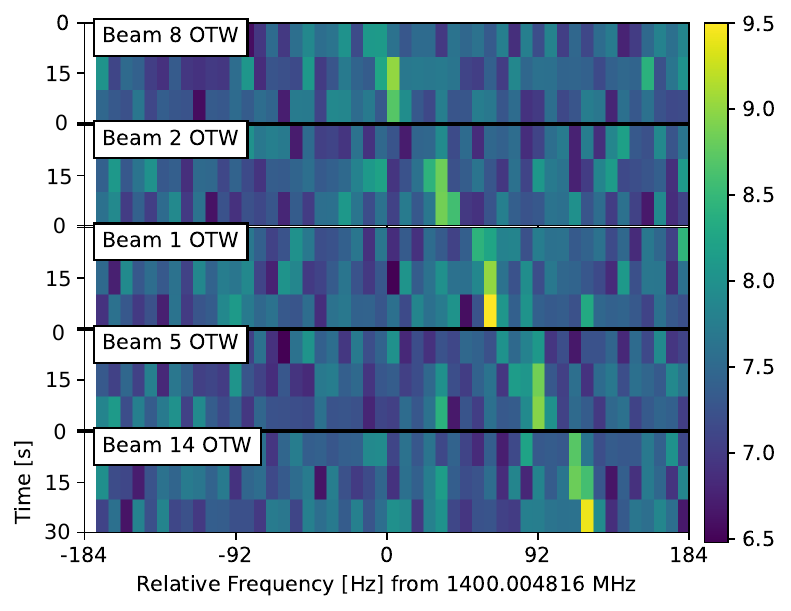}
}
\subfigure[]{
\includegraphics[width=0.4\linewidth]{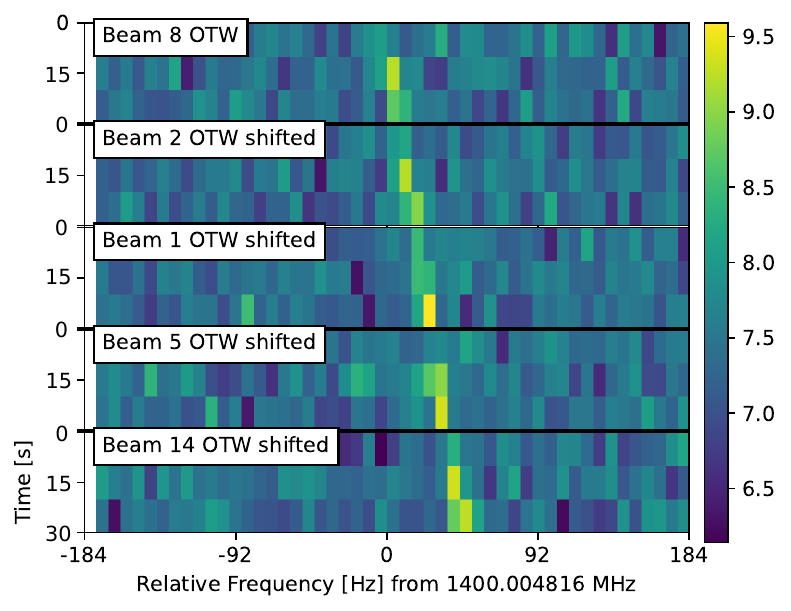}
}

\caption{Waterfall plots of the OTBs from a simulated MBPS observation with a slew rate of 6 arcsec/s. The colored bar indicates the power with arbitrary units. (a) waterfall plot of a high SNR signal in the five OTBs during the whole observation. (b) same as (a) but with lower SNR. (c) waterfall plot of the lower SNR signal in the five OTWs. (d) same as (c) but the central frequencies of beams 2, 1, 5, and 14 are subtracted by the frequency drifts during the corresponding gaps between OTWs. }
\label{fig:6arcsec}
\end{figure}

In the re-observation of TRAPPIST-1, the slew rate is kept constant at 1 arcsec/s, which makes an OTW length of 181 seconds as shown in Table. \ref{tab:otw}. With a time resolution of 10 seconds, a narrowband ($\sim$ Hz) ETI signal would span approximately 18 pixels in a single OTW in the time axis. If an ETI signal persists for less than 181 seconds it means the signal is possible to be visible only in one OTW and exhibits intensity variation as mentioned in the section of cross-verification for ETI signals, though the MBPS strategy will not mislabel it as RFI, since there are no other OTWs to cross-verify, we are less confident to confirm it as ETI signal and required some extra verification processes such as comparing the frequency and morphology with know RFI, etc. However, a faster slew rate may reduce the minimum ETI signal duration for cross-verification. A future observation could increase the slew rate up to 6 arcsec/s which means an OTW length of 30 seconds, thus a minimum ETI signal duration of 30 seconds can be cross-verified with different OTWs. Nonetheless, there are obvious drawbacks to a faster slew rate, with an OTW length of 30 seconds, there would be only three pixels in the time axis to represent the signal which increases the probability that the three pixels are recognized as noise points by the search algorithm, and it is even worse for weaker signals as shown in Fig. \ref{fig:6arcsec}(a)(b). A possible solution could be done before running the observation data on signal-searching algorithms which is to segment each beam's OTW, then splice them together to be examined as shown in Fig. \ref{fig:6arcsec}(c). Again, all we need to do now is to correct the frequencies for the drifting during the gaps between OTWs. Fig. \ref{fig:6arcsec}(d) shows the complete signal after correcting the frequencies for beams 2, 1, 8, and 14, now we have connected five of 3 pixels, sum up to 15 pixels in the time axis which is sufficient for the search algorithm to recognize as a line with proper slope.

\subsection{Applications of the non-OTBs}
Non-OTBs are the beams that never point at the target star during the observation. Although beyond the scope of this paper, studying the non-OTBs using the MBPS strategy allows us to effectively expand our surveyed sky area and potentially detect ETI signals from the background astrophysical objects.
\section{Conclusion and Discussion} \label{sec: CD}

Results from the re-observation of TRAPPIST-1 using the MBPS strategy match precisely with our predictions. TurboSETI has found 6972 hits. With the MBPS, we eliminate all hits from being potential ETI signals, because most of them are visible throughout the entire observation in a single beam. There are two major advancements achieved by the MBPS strategy in terms of identifying continuous RFI with certainty and confirming ETI signals by cross-verification with different beams in a single observation which essentially enables the MBPS strategy to eliminate nearly all false positives and false negatives. The way that the MBPS strategy achieves these is that we are effectively adding new parameters and the observation data can thus be interpreted from different perspectives. The additional parameters introduced by the MBPS strategy include the duration of signals in a single beam, intensity variation of signals, and the difference in central frequencies of different beams which are the results of the observation method of the MBPS. With the three newly introduced parameters, we are then able to put in the most rigorous restrictions on the RFI/ETI identifications by confining the characteristics of an ETI/RFI signal in a new multi-parameter space. We speculate that it would be exceedingly rare for the MBPS strategy to return any anomalous signals, even over the course of several years, because the types of false positives found by other strategies are easily identifiable with the MBPS strategy. However, when a genuine narrowband ETI signal does arrive on Earth, the MBPS strategy is capable of identifying it even amidst a substantial influx of RFI.

\vspace{\baselineskip}

In the ON-OFF strategy, signals that are present in all 'ON' observations but not in any of the 'OFF' observations are called events. Events that are located within certain known RFI concentrated frequency bins will be marked as RFI without actual evidence which means an ETI signal that is sent coincidentally close to those known RFI frequency regions will be marked as RFI. However, the MBPS strategy identifies continuous RFI with solid reasoning by examining the new parameters which can not be achieved by any previous strategies.

\vspace{\baselineskip}

Furthermore, signals that pass the RFI identification pipeline of previous strategies and are similar in morphology to expected ETI signals located at a particular frequency that is not near any known RFI-contaminated regions, will be considered as candidate ETI signals and required extra analysis or re-observation to confirm as in the ON-OFF and MBCM strategies. MBPS, however, is capable of identifying such signals by inspecting their duration, time window, intensity variation, and finally the cross-verification with different beams which requires no further observation for a reliable RFI or ETI confirmation.

\vspace{\baselineskip}

Recent SETI developments have been accelerated mainly by data analysis techniques like machine learning \citep{pinchuk2022machine,zhang2020first,ma2023deep} However, while we acknowledge the significance of data analysis, it is imperative to avoid confining our development solely to it. Data analysis techniques facilitate a rapid comprehension of the information contained within a dataset. Nevertheless, employing a targeted data collection approach has the potential to expand the maximum extent of information encompassed by the data. The method introduced in this work aims to do so by exploring a different observation strategy. Moreover, since we put a more complete and strict definition on a truly sky-localized signal by introducing more parameters, we are potentially reducing the difficulty for a future deep-learning neural network to learn the particular features of RFI or synthetic ETI signals in the training process.

\section*{Acknowledgments}

The authors would like to thank the referees for their constructive suggestions, which greatly contributed to the enhancement of this work. We thank Jing-Hai Sun for providing the configuration of the FAST's 19-beam receiver. This work was supported by the National SKA Program of China (2022SKA0110202) and the National Natural Science Foundation of China (grants No. 11929301). The data analysis is done on the workstation at Dezhou University.



\bibliography{text}



\end{document}